\documentclass{article}
\usepackage{PRIMEarxiv}
\usepackage{multirow}
\usepackage{subfig}
\usepackage{balance}
\usepackage[utf8]{inputenc} 
\usepackage[T1]{fontenc}    
\usepackage{url}            
\usepackage{booktabs}       
\usepackage{amsfonts}       
\usepackage{nicefrac}       
\usepackage{microtype}      
\usepackage{lipsum}
\usepackage{fancyhdr}       
\usepackage{graphicx}       
\usepackage{amsmath}
\usepackage{tabularx}
\usepackage{array}
\usepackage{bbding}
\usepackage{amssymb}
\usepackage{multicol}
\usepackage{multirow}
\usepackage{mwe}
\usepackage{pdfpages}
\usepackage{float}
\usepackage{hyperref}
\usepackage{amssymb}
\usepackage{algorithm}
\usepackage{algorithmicx}
\usepackage{algpseudocode}
\usepackage{color}
\usepackage{xcolor}

\author{
    Xueru Wen \\
    University of Chinese Academy of Sciences \\
    Beijing, China \\
    \texttt{wenxueru2022@iscas.ac.cn} \\
\And
    Xiaoyang Chen \\
    University of Chinese Academy of Sciences \\
    Beijing, China \\
    \texttt{chenxiaoyang19@mails.ucas.ac.cn}
\And
    Xuanang Chen \\
    University of Chinese Academy of Sciences \\
    Beijing, China \\
    \texttt{chenxuanang19@mails.ucas.ac.cn}
\And
    Ben He\thanks{*Corresponding author}\\
    University of Chinese Academy of Sciences \\
    Beijing, China \\
    \texttt{benhe@ucas.ac.cn} \\
\And
    Le Sun \\
    Institute of Software, Chinese Academy of Sciences \\
    Beijing, China \\
    \texttt{sunle@iscas.ac.cn} \\
}

\title{Offline Pseudo Relevance Feedback for Efficient and Effective Single-pass Dense Retrieval}

\begin{document}
\maketitle

\begin{abstract}
Dense retrieval has made significant advancements in information retrieval (IR) by achieving high levels of effectiveness while maintaining online efficiency during a single-pass retrieval process. 
However, the application of pseudo relevance feedback (PRF) to further enhance retrieval effectiveness results in a doubling of online latency. 
To address this challenge, this paper presents a single-pass dense retrieval framework that shifts the PRF process offline through the utilization of pre-generated pseudo-queries. 
As a result, online retrieval is reduced to a single matching with the pseudo-queries, hence providing faster online retrieval. 
The effectiveness of the proposed approach is evaluated on the standard TREC DL and HARD datasets, and the results demonstrate its promise. Our code is openly available at \href{https://github.com/Rosenberg37/OPRF}{https://github.com/Rosenberg37/OPRF}.
\end{abstract}


\twocolumn

\section{Introduction}

The advancement of dense retrieval with pre-trained language models \cite{yates-etal-2021-pretrained} has significantly improved the benchmark due to their remarkable text representation ability. 
These models \cite{10.1145/3397271.3401075, karpukhin-etal-2020-dense, https://doi.org/10.48550/arxiv.2007.00808} typically pre-compute the document representation and perform efficient single-pass retrieval by applying relevance weighting. 
Moreover, Pseudo Relevance Feedback (PRF) has been extensively explored to enhance the retrieval performance in dense retrieval. 
PRF utilizes the top-ranked documents to update the query representation, followed by a second-pass retrieval through the revised query \cite{https://doi.org/10.48550/arxiv.2108.11044, Rocchio1971RelevanceFI, lv_comparative_2009}. 
Despite its effectiveness, as reported in \cite{li-etal-2018-nprf, 10.1007/978-3-030-72240-1_46, https://doi.org/10.48550/arxiv.2108.11044}, the employment of PRF introduces the extra computational cost of dense retrieval, particularly for large collections where the online latency grows linearly with the collection size.

In this paper, a novel single-pass dense retrieval method is presented to reduce the computation required for the PRF and the subsequent second-pass dense retrieval by shifting a substantial portion of these computations to the offline stage. 
During the offline stage, dense PRF is conducted for a set of pre-generated pseudo-queries for the target documents. The top-ranked results for each of these pseudo-queries are then stored for future use.
In the online retrieval stage, we apply the lightweight lexical matching function BM25 \cite{Robertson1995OkapiAT} to search for the pseudo-queries for the replacement of the original query.
Once locating the relevant pseudo-queries, their pre-stored results are aggregated to produce the final ranking list. 
This reduces the online retrieval to a simple sparse retrieval process that only requires unsupervised lexical matching between the user query and the short pseudo-queries. 
Moreover, compared to the previous offline matching approach \cite{10.1145/3404835.3463073}, our proposed method has lower computational overhead in both offline and online stages. 

The contributions of this work are tri-fold: 1) Proposal of a single-pass dense retrieval framework transferring the heavy-weight computation of pre-trained language model (PLM) and PRF procedure to the offline. 2) Integrating different dense retrieval methods with PRF techniques while maintaining low latency. 3) Comprehensive experiments on benchmarks showcase its effectiveness. 

\section{Related Work}
\noindent\textbf{Dense retrieval.} The retrieval models based on PLMs \cite{https://doi.org/10.48550/arxiv.2211.14876} have been developed into a new trend of information retrieval.
There are two typical architectures for dense retrieval, viz. cross-encoder and bi-encoder.
The cross-encoder architecture \cite{Qiao2019UnderstandingTB, 10.1145/2983323.2983769, DBLP:conf/sigir/XiongDCLP17} takes the concatenation of the query and the document as input and utilizes PLMs to catch the semantic interaction.
The bi-encoder \cite{karpukhin-etal-2020-dense, qu-etal-2021-rocketqa, https://doi.org/10.48550/arxiv.2007.00808} architecture adopts the two-tower architecture and transforms query and document into single- or multi- representation \cite{karpukhin-etal-2020-dense,10.1145/3397271.3401075} allowing the pre-computation of document representation for efficiency.

\noindent\textbf{Pseudo relevance feedback} was first proposed to ease the term mismatch problem \cite{10.1145/32206.32212} and now increasingly raises attention due to its effectiveness in dense retrieval.
Many \cite{li-etal-2018-nprf, 10.1016/j.ipm.2019.102182, 10.1007/978-3-030-72240-1_46} have put efforts to employ PRF techniques in dense retrieval and reach a satisfactory result. 
Though developed, current PRF approaches may be impeded by the increased computational requirements, which could restrict their widespread use for very large datasets.

\noindent\textbf{PLM-based reranker.} Pre-trained language model \cite{devlin-etal-2019-bert} has been explored to construct re-ranker \cite{DBLP:journals/corr/abs-1901-04085, DBLP:journals/corr/abs-1905-09217} to further re-order the candidate passages returned by the first-pass retriever.
The cross-encoder architecture is typically employed for the re-ranker \cite{Qiao2019UnderstandingTB, DBLP:conf/trec/YanLWBWXS19, DBLP:conf/emnlp/WangNMNX19} with substantial improvement while burdening the online latency.
The contextualized relevance offline weighting (CROW) \cite{10.1145/3404835.3463073} approach is proposed following the idea of splitting the pipeline into the online and the offline stages \cite{10.1145/2348283.2348368, 10.1145/3345001}.
Despite the gain in efficiency, the heavy offline computational cost hinders further enhancing effectiveness with techniques like PRF, which is attempted to address in this paper through lightweight integration of PRF.

\section{Method}

\subsection{Overview}
In the proposed framework, relevance weighting is divided into online and offline stages. 
In the offline stage, the relevance scores between the pseudo-queries and documents are computed. 
In the online stage, BM25 \cite{Robertson1995OkapiAT} is used to identify the appropriate pseudo-queries to replace the original query and then combine it with the pre-computed relevance scores to produce the final ranking.

\subsection{Offline Preparation}
The offline preparation aims to locate the documents matching the different search intents expressed by pseudo-queries pre-generated for each document.
Given a corpus $\mathcal{C}$, $m$ pseudo-queries are generated for each document $d \in \mathcal{C}$ through a seq2seq model \cite{Cheriton2019FromDT, Nogueira2019DocumentEB}, denoted as $\mathcal{Q}_d = \{\overline{q}_1,\overline{q}_2,...,\overline{q}_m\}$. The final pseudo-queries set $\mathcal{Q}$ are the union of all pseudo-queries sets, i.e. $\mathcal{Q} = \cup_{d \in \mathcal{C}} \mathcal{Q}_d$. 

For each $\overline{q} \in \mathcal{Q}$, a list of top-\textit{k} documents, denoted as $\mathcal{D}_{\overline{q}}$, are returned by designated dense retrieval model with the support of PRF enhancement and their relevance scores will be stored for further use during the online retrieval. 
As the laborious retrieval computations and deployment of PRF have been done offline, our framework achieves the reduction of online latency.

\subsection{Online Retrieval}
As the dense retrieval with PRF has been transferred offline, online retrieval mainly involves a simple lexical matching process. 
Specifically, given a query $q$ for online retrieval, BM25 \cite{Robertson1995OkapiAT} is first utilized to fetch the top-$s$ pseudo-queries, denoted as $\mathcal{S}_q=\{\overline{q}_1,\overline{q}_1,...,\overline{q}_s\}$.
The final set of candidate documents $\mathcal{R}_q$, is the union of all the neighboring document sets $\mathcal{D}_{\overline{q}}$, thus a collection containing $r$ documents, where $r \le s\times k$ as not all documents in $\mathcal{R}_q$ have relevance scores assigned to every pseudo-query in $\mathcal{S}_q$. 

Once obtain the candidate document set $\mathcal{R}_q$, we proceed to compute the final relevance scores for each document $d \in \mathcal{R}_q$. 
We use two similarity scores: one between the original query and the pseudo-query, represented as $sim\left(q,\overline{q}\right)$, is given as the BM25 score as previously described; and another between the pseudo-query and the document, represented as $sim\left(\overline{q}, d\right)$, is pre-computed offline by the selected dense retrieval model.
We apply min-max normalization to the $sim\left(\overline{q}, d\right)$ along the pseudo-queries as follows:
\begin{equation}
\resizebox{\hsize}{!}{$
     {sim}^{\prime}\left(\overline{q}, d\right)=\frac{sim\left(\overline{q}, d\right) - {min}_{d_i \in \mathcal{R}_q}\left(sim\left(\overline{q}, d_i\right)\right)}{{max}_{d_i \in \mathcal{R}_q}\left(sim\left(\overline{q}, d_i\right)\right) - {min}_{d_i \in \mathcal{R}_q}\left(sim\left(\overline{q}, d_i\right)\right)}$}
\end{equation}
The normalization step ensures the relevance scores are on the same scale. 
To produce the final relevance score as a scalar, we first apply a softmax operation to the relevance scores between the original query and the pseudo-queries, transforming $sim\left(q,\overline{q} \right)$ into weighted terms and then rank documents $d \in \mathcal{R}_q$ by a weighted sum of scores of different pseudo-queries ${sim}^{\prime}\left(\overline{q}, d\right)$ as follows:
\begin{equation}
     {sim}\left(q,d\right)=\sum_{\overline{q} \in \mathcal{S}_q} \frac{e^{sim\left(q,\overline{q} \right)}}{\sum_{\overline{q}_i \in \mathcal{S}_q} e^{sim\left(q,\overline{q}_i \right)}} \times {sim}^{\prime}\left(\overline{q},d\right)
\end{equation}


\subsection{Score Fusion}

\textbf{Fusion of different models}. We follow \cite{10.1145/3477495.3531884} and extend our framework by fusing multiple models with or without PRF through weighted sum by the softmax normalized terms as follows:
\begin{equation}
         {sim}\left(q,d\right)=\sum_{\overline{q} \in \mathcal{S}_q, a \in \mathcal{A}} \frac{e^{sim\left(q,\overline{q} \right)}}{\sum_{\overline{q}_i \in \mathcal{S}_q} e^{sim\left(q,\overline{q}_i \right)}} \times {sim}^{\prime}_a\left(\overline{q},d\right)
\end{equation}
\noindent where $\mathcal{A}$ denotes the set of retrieval models and ${sim}_a\left(\overline{q},d\right)$ represent the similarity score given by the model $a$. 
This extension shows prominent results without a significant impact on online latency and is hence used as the default setting of our experiments.

\noindent\textbf{Fusion with the original query}. Another variant is to utilize the original query $q$ to avoid potential semantic shifts caused by pseudo-query. 
Specifically, the original query $q$ can be seen as a pseudo-query, and its similarity score $sim\left(q,q\right)$ is set to the maximum score of $sim\left(q,\overline{q}\right)$. 
Since the result of the original query cannot be prepared in advance, the efficient sparse algorithm is used for retrieval, and results are then aggregated with other pseudo-queries.

\begin{table*}[tbh]
\caption{Retrieval results of passage ranking tasks. Statistical significant differences relative to docT5query and uniCOIL + docT5query at $p < 0.05$ are denoted as $\ast$ and $\star$, respectively. The results of models with and without PRF are separated by a slash. $\dagger$ denotes the fusion with retrieval results of the original query.}
\label{tab:trec}
\resizebox{\linewidth}{!}
{
\begin{tabular}{lcccccccccccc}
\hline
\multicolumn{1}{c}{\multirow{2}{*}{Model/PRF}} & \multirow{2}{*}{Latency}    & \multicolumn{3}{c}{TREC DL 2019}                                                                                           &  & \multicolumn{3}{c}{TREC DL 2020}                                                                   &  & \multicolumn{3}{c}{DL-HARD}                                                                                  \\ \cline{3-5} \cline{7-9} \cline{11-13} 
\multicolumn{1}{c}{}                           &                             & nDCG@10                       & MAP                                              & R@1k                                    &  & nDCG@10                       & MAP                                & R@1k                          &  & nDCG@10              & MAP                                         & R@1k                                    \\ \hline
BM25/RM3                                       & $1\times$/$2.7\times$       & 0.506/0.522                   & 0.301/0.342                                      & 0.750/0.814                             &  & 0.480/0.482                   & 0.286/0.302                        & 0.786/0.822                   &  & 0.290/0.264          & 0.164/0.154                                 & 0.678/0.699                             \\
docT5query                                     & $1.5\times$                 & 0.634                         & 0.405                                            & 0.813                                   &  & 0.627                         & 0.417                              & 0.839                         &  & 0.366                & 0.217                                       & 0.772                                   \\
uniCOIL+docT5query                           & $27.9\times$                & 0.703                         & 0.462                                            & 0.829                                   &  & 0.675                         & 0.443                              & 0.843                         &  & 0.360                & 0.209                                       & 0.769                                   \\ \hline
CROW                                           & $8.4\times$                 & 0.692                         & 0.510                                            & 0.816                                   &  & 0.664                         & 0.453                              & 0.861                         &  & -                    & -                                           & -                                       \\
ANCE/Avg                                       & $48.7\times$/$73.5\times$   & 0.645/0.653                   & 0.371/0.425                                      & 0.755/0.774                             &  & 0.646/0.657                   & 0.408/0.436                        & 0.776/0.791                   &  & 0.334/0.323          & 0.195/0.200                                 & 0.767/0.734                             \\
DBERT KD TASB/Avg                         & $55.4\times$/$84.1\times$   & 0.721/0.719                   & 0.459/0.486                                      & 0.841/0.852                             &  & 0.685/0.709                   & 0.470/0.489                        & 0.873/0.903                   &  & 0.376/\textbf{0.391} & 0.238/0.241                                 & 0.827/0.809                             \\
TCT-ColBERT/Avg                                & $42.2\times$/$87.8\times$   & 0.670/0.664                   & 0.391/0.434                                      & 0.792/0.823                             &  & 0.668/0.696                   & 0.429/0.473                        & 0.818/0.867                   &  & 0.369/0.355          & 0.223/0.213                                 & 0.810/0.812                             \\
TCT-ColBERTv2/Avg                              & $36.7\times$/$74.1\times$   & 0.720/0.711                   & 0.447/0.488                                      & 0.826/0.859                             &  & 0.688/0.684                   & 0.475/0.481                        & 0.843/0.858                   &  & 0.372/0.353          & 0.224/0.215                                 & 0.801/0.785                             \\
Fusion/Avg                                     & $157.2\times$/$243.0\times$ & 0.734/\textbf{0.748}          & 0.480/0.518                                      & 0.863/\textbf{0.905}                    &  & 0.722/\textbf{0.725}          & 0.501/\textbf{0.520}               & 0.879/\textbf{0.910}          &  & 0.380/0.385          & 0.240/0.249                                 & \textbf{0.836}/0.831                    \\ \hline
Ours/Avg                                       & $2.6\times$/$2.5\times$     & $0.713^{\ast}$/$0.711^{\ast}$ & $0.470^{\ast}$/$0.506^{\ast}$                    & $0.870^{\ast}$/$0.894^{\ast\star}$      &  & $0.677^{\ast}$/$0.694^{\ast}$ & $0.469^{\ast}$/$0.495^{\ast\star}$ & 0.866/$0.893^{\ast}$          &  & 0.372/0.365          & 0.232/0.247                                 & 0.797/0.802                             \\
$\text{Ours}^{\dagger}$/Avg                    & $5.2\times$/$4.9\times$     & $0.728^{\ast}$/$0.729^{\ast}$ & $0.517^{\ast\star}$/$\textbf{0.531}^{\ast\star}$ & $0.891^{\ast\star}$/$0.891^{\ast\star}$ &  & $0.687^{\ast}$/$0.696^{\ast}$ & $0.476^{\ast}$/$0.502^{\ast\star}$ & $0.881^{\ast}$/$0.881^{\ast}$ &  & 0.388/0.387          & $0.253^{\ast}$/$\textbf{0.267}^{\ast\star}$ & $0.824^{\ast\star}$/$0.824^{\ast\star}$ \\ \hline
\end{tabular}
}
\end{table*}

\section{Experiments}
\subsection{Experiment Settings}

\textbf{Dataset and metrics.} MS MARCO \cite{nguyen2016ms} is used in our experiments.
The queries from TREC 2019 and 2020 Deep Learning (DL) Track \cite{https://doi.org/10.48550/arxiv.2003.07820, https://doi.org/10.48550/arxiv.2102.07662} and the more complex topics from the Deep Learning Hard (DL-HARD) dataset \cite{DBLP:journals/corr/abs-2105-07975} are utilized. We evaluate the effectiveness in terms of nDCG@10, MAP, and R(recall)@1k and measure the efficiency by online latency excluding the time of pre-processing and reading indexes or other data into memory. Statistical significance at $p < 0.05$ according to the paired two-tailed t-test is reported.

\textbf{Competing methods.} We compare our approach with several strong baselines, covering a wide range of sparse and dense retrieval techniques. 
\textbf{BM25} \cite{Robertson1995OkapiAT} is a classical unsupervised ranking model based on lexical matching, on top of which \textbf{BM25+RM3} \cite{10.1145/383952.383972} applies the RM3 relevance feedback. 
\textbf{docT5query} \cite{Cheriton2019FromDT} expands documents by pseudo-queries generated by a pre-trained T5 model to mitigate the term mismatch problem. 
\textbf{uniCOIL + docT5query} \cite{10.1145/3404835.3462891, gao-etal-2021-coil} performs exact matching over dense embeddings on the expanded documents. 
\textbf{TCT-ColBERT} \cite{https://doi.org/10.48550/arxiv.2010.11386} and \textbf{TCT-ColBERTv2} \cite{lin-etal-2021-batch} distill the ColBERT late-interaction model \cite{10.1145/3397271.3401075} into efficient dense retrieval models. 
\textbf{ANCE} \cite{https://doi.org/10.48550/arxiv.2007.00808} filters hard training negatives from the corpus for training, and \textbf{DistilBERT KD TASB} \cite{10.1145/3404835.3462891} adapts knowledge distillation to the different ranking architectures. 
\textbf{CROW} \cite{10.1145/3404835.3463073} decompose the relevance weighting into an offline stage and online stage and shift the heavy computation offline.
The aforementioned models are reproduced by Pyserini \cite{10.1145/3404835.3463238} with default parameter settings, except that the results of CROW reported in \cite{10.1145/3404835.3463073} are compared. 
The dense models with the vector-based PRF on average \cite{https://doi.org/10.48550/arxiv.2108.11044} of top-$3$ feedback passages, denoted as \textbf{Avg}, is included.
\textbf{Fusion} represents fusing different retrieval models directly applied on the user query and \textbf{Fusion-Avg} denotes further deploying the PRF method.
The variants with or without PRF and with or without fusing the original query of our approach are evaluated.
Three models are adopted to form the offline system, viz. TCT-ColBERTv2, DistilBERT KD TASB, and uniCOIL + docT5query.
The sparse algorithm for fusing the original query is BM25 + RM3 with docT5query expansion.

\textbf{Implementation details}. In the online stage, each user query is substituted by $s=4$ pseudo-queries, and each pseudo-query stores $k=1000$ top-ranked documents. During the offline preparation, we use the pseudo-queries generated by docT5query \cite{Cheriton2019FromDT}, where each document is associated with $m=80$ pseudo-queries. 

\subsection{Results} 
According to Table \ref{tab:trec}, \textbf{comparing to the sparse retrieval baselines}, all variants of our framework provide remarkable improvement in effectiveness while maintaining efficiency.
The latency of our method is slightly higher than the lexical matching algorithms due to the larger inverted index and the extra aggregation operation, but lower compared to uniCOIL + docT5query by virtue of the avoidance of online feed-forward of the PLM.
Through the performance profiling, we discover that the pseudo-query matching constitutes roughly half of the online latency and extra operations, viz. aggregation, normalization, and read caching account for the other half.
We also report the significant differences with respect to docT5query and uniCOIL + docT5query, both of which is the same as us involving only exact matching during the online stage, and our method gains significant improvement in most metrics.

In \textbf{comparison with the dense retrieval baselines}, our best results outperform the individual models forming the offline system, and all variants reach comparable results. 
Moreover, our method obtains the consistent increase brought by PRF without latency deterioration in contrast to the other approaches.
In \textbf{competing with the Fusion(-Avg) baseline}, our approach is marginally lower due to the semantic shift of pseudo-queries. 
But these advancements are under markedly grown online latency, losing the efficiency of our approach. 
Also, fusion with the original query is conducive to alleviating the effectiveness drop caused by pseudo-queries.

\subsection{Analysis}
\begin{figure*}[tbp]
    \centering
    \subfloat[\# Pseudo-queries returned online.]{
        \label{fig:num_pseudo}
        \includegraphics[width=0.32\linewidth]{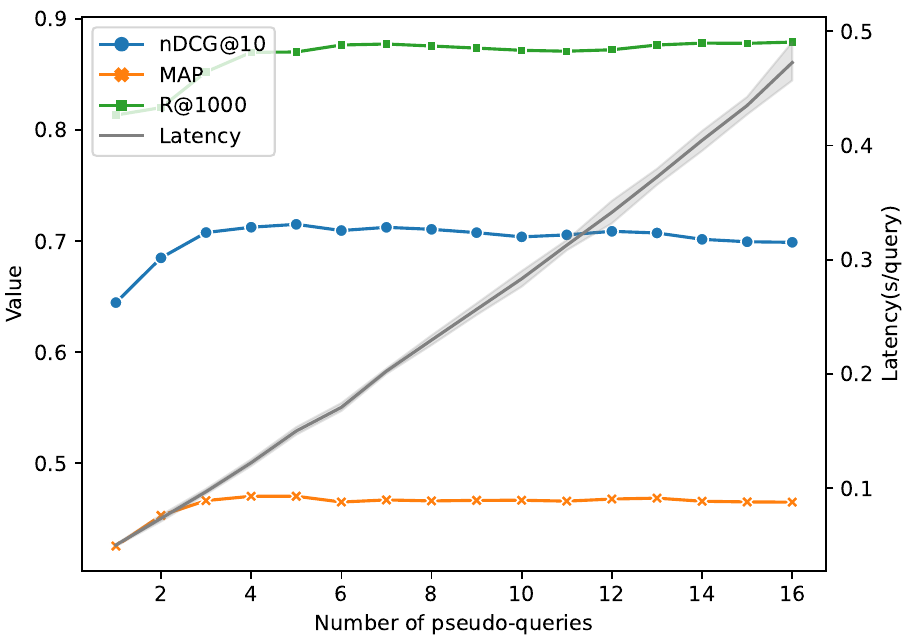}
    }
    \subfloat[\# Documents retrieved per pseudo-query offline.]{
        \label{fig:num_return}
    	\includegraphics[width=0.32\linewidth]{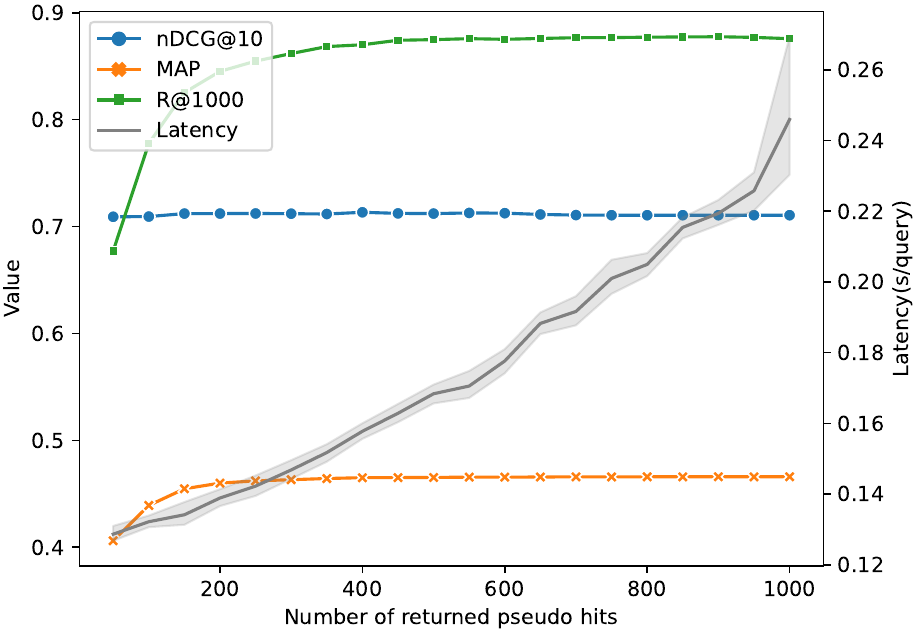}
    }
    \subfloat[\# Pseudo-queries generated per document.]{
        \label{fig:num_generate}
    	\includegraphics[width=0.32\linewidth]{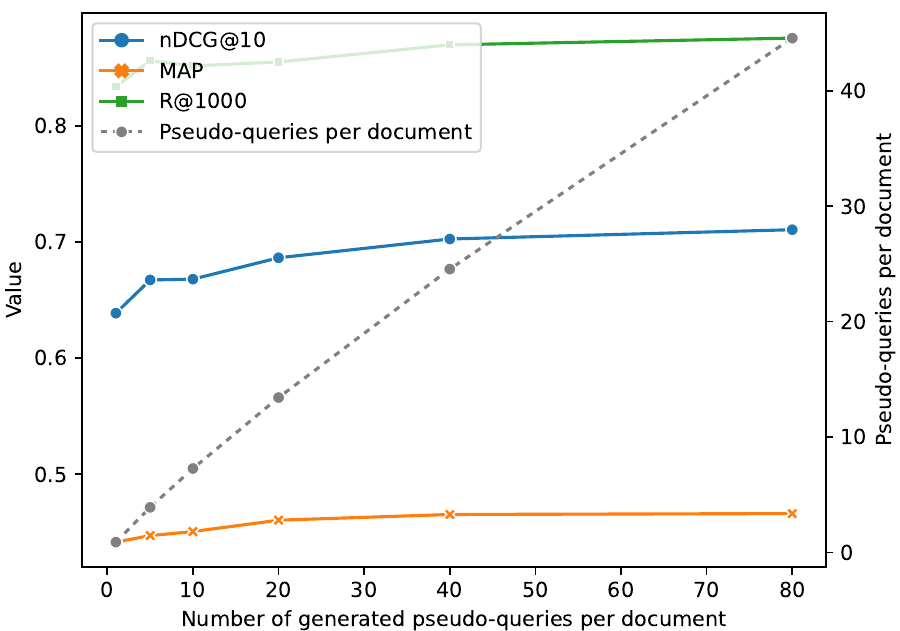}
    }
    \caption{Influence of different hyper-parameters on TREC 2019 DL Passage Ranking Track.}
    \label{fig:hyper}
\end{figure*}

\subsubsection{Influence of hyperparameters}
The impact of \textbf{the number of pseudo-queries returned online} is depicted in Figure \ref{fig:num_pseudo}. 
As the hyper-parameter increases from $1$ to approximately $4$, improvements are observed, which can be attributed to an increased number of high-quality pseudo-queries better capturing the semantics of the user query.
Beyond that, more pseudo-queries do not seem very beneficial, indicating the marginal contribution of low-quality pseudo-queries. 
Furthermore, the online latency rises with the hyper-parameter value, due to a larger number of pre-computed results to be aggregated.
Despite this, we note that the lag is still smaller compared to the dense baselines, as shown in Table \ref{tab:trec}.

\textbf{The number of documents retrieved by each pseudo-query offline}, as depicted in Figure \ref{fig:num_return}, exhibits behavior where MAP and R@1k initially increase and then plateau, while nDCG@10 evaluating the top-$10$ results does not show significant fluctuations. 
It can be attributed to the decline in relevance scores as the ranking of the documents decreases, thereby reducing the likelihood of their inclusion in the final set of results. 
Additionally, as more results are aggregated, the latency increases, as demonstrated in Figure \ref{fig:num_return}.

The impact of \textbf{the number of pseudo-queries generated per document} is examined in Figure \ref{fig:num_generate}. 
The figure indicates the limit to the enhancement of effectiveness through expanding the pseudo-queries set.
$5$ pseudo-queries per document can achieve a sufficient level of retrieval effectiveness. 
Additionally, a significant proportion of the pseudo-queries are duplicated as only approximately 60\% pseudo-queries are left after the union of pseudo-queries sets.

\subsubsection{Online efficiency}
\begin{figure}
  \centering
  \includegraphics[width=\linewidth]{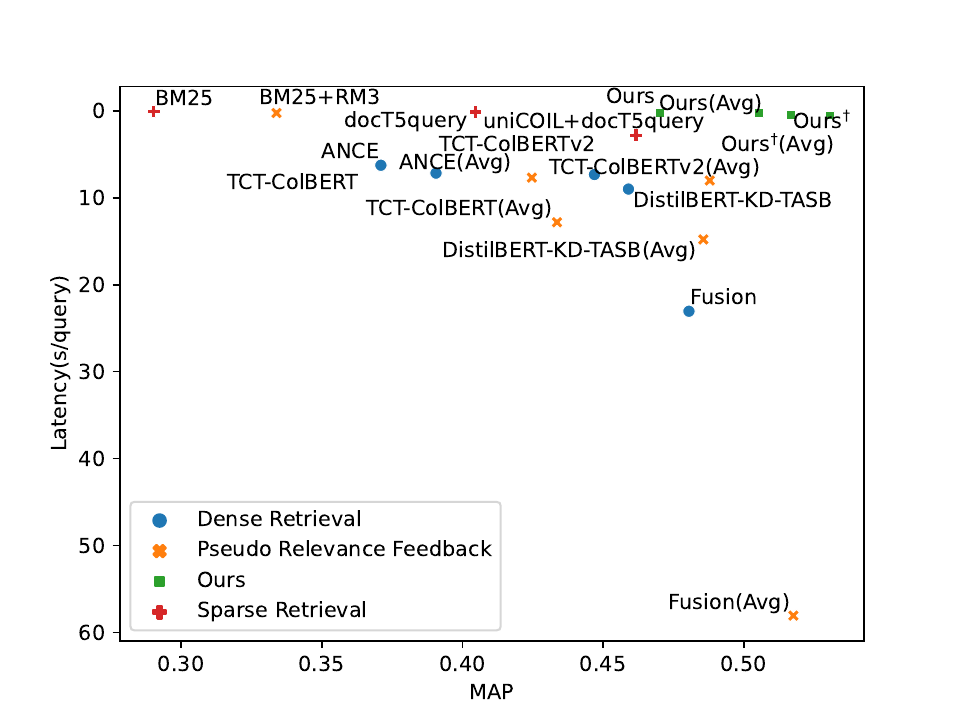}
  \caption{Effectiveness (MAP) versus per-query latency on TREC 2019 DL Passage Ranking task.}
  \label{fig:efficiency}
\end{figure}

Figure \ref{fig:efficiency} demonstrates the online latency and effectiveness of different models. 
Our proposed approach performs with a high level of efficiency, comparable to the sparse models such as BM25 \cite{Robertson1995OkapiAT}. 
The utilization of the strengths of dense retrieval and PRF in the offline process significantly enhances effectiveness in MAP. 
The superiority of our approach in terms of reduced online latency is readily apparent compared to the dense retrieval baselines. 
It is because our approach involves only a simple lexical matching between queries, which typically consist of short texts. 
A distinct advantage of our approach in contrast to the existing dense retrieval paradigm is that the online latency of our approach does not increase significantly with collection size.
We do not plot for nDCG@10 and R@1k as the results are correlated.

\subsubsection{Offline cost}
As for the disk footprint, our approach shares the same composition as CROW \cite{10.1145/3404835.3463073}, storing pseudo-queries, relevance scores, and document IDs, but requires less space due to the deduplication under the identical hyperparameter setting of pseudo-queries generated per document and documents retrieved per pseudo-query.
Supposing that generate $80$ pseudo-queries per document, one document corresponds roughly to $45$ in pseudo-queries according to Figure \ref{fig:num_generate}.
Each pseudo-query has approximately $34$ letters ($1$ byte/letter) and stores $1000$ relevance scores ($4$ bytes/score) and document IDs ($4$ bytes/ID). 
Consequently, each document takes $361530$ bytes ($3.53\times$).
Assuming each document on average has $200$ tokens and is represented in $128$-dimensional vectors ($4$ bytes/dimension), one document takes $102400$ bytes ($1\times$) in ColBERT \cite{10.1145/3397271.3401075}. 
If we generate a smaller set of pseudo-queries, e.g. generating only $5$ pseudo-queries per document and recalling $500$ documents per pseudo-query, the storage cost per document is substantially decreased to $15,733$ ($0.15\times$) bytes. 
Note that this tradeoff only has a slight impact on effectiveness, as shown in Figure \ref{fig:hyper}. Our approach still achieves MAP scores of $0.453$ and $0.500$ with and without PRF, respectively, which is comparable to the baselines.


As for computational cost, according to our estimation, it roughly takes $188$ days to prepare the relevance scores between pseudo-queries and documents for the entire MS Marco passage corpus by TCT-ColBERTv2 on one Titan RTX 24G GPU and one Intel E5-2680 v4 CPU running at 2.40GHz. 
This is only 3\% of the offline computation cost of CROW \cite{10.1145/3404835.3463073}, which relies on a heavy cross-encoder for offline matching. Note that the offline computation of our approach can be easily scaled up by various means, e.g. parallel computing with more GPU/CPU threads, ANNS algorithms \cite{10.5555/2283516.2283615, 8681160}, and adopting TPUs. Additionally, it is actually unnecessary to pre-calculate scores for all pseudo-query-document pairs and a grow-with-search score cache is another feasible improvement.

\section{Conclusions}
This paper presents a single-pass dense retrieval framework shifting the pseudo relevance feedback (PRF) process to the offline through pre-generating pseudo-queries and separating the computation of relevance scores into online and offline stages. 
We thoroughly evaluate our framework using the TREC DL and HARD datasets and showcase its ability to achieve both efficient and effective dense retrieval with PRF. 
In future research, we aim to explore the integration of more dense retrieval models and investigate the application of more advanced generative language models for offline pseudo-queries generation. 
This approach has the potential to improve the identification of diverse search intents within a document, ultimately leading to more precise and relevant retrieval results.

\section{Acknowledgments}
This research work is supported by the National Natural Science Foundation of China under Grants no. 62272439, and the Project of the Chinese Language Committee under Grant no. YB2003C002.

\bibliographystyle{unsrt}  
\bibliography{references}  

\end{document}